# Single-step fabrication of high performance extraordinary transmission plasmonic metasurfaces employing ultrafast lasers

Carlota Ruiz de Galarreta[1*,2‡], Noemi Casquero[1‡], Euan Humphreys[2], Jacopo Bertolotti[2], Javier Solis[1], C. David Wright[2], and Jan Siegel[1*]

1 Laser Processing Group, Instituto de Óptica, IO-CSIC, Serrano 121, Madrid, 28006 Spain

2 College of Engineering, Mathematics and Physical Sciences, University of Exeter, Exeter, EX4 4QF UK.

**ABSTRACT:** Plasmonic metasurfaces based on the extraordinary optical transmission (EOT) effect can be designed to efficiently transmit specific spectral bands from the visible to the far-infrared regimes, offering numerous applications in important technological fields such as compact multispectral imaging, biological and chemical sensing, or color displays. However, due to their subwavelength nature, EOT metasurfaces are nowadays fabricated with nano- and micro-lithographic techniques, requiring many processing steps and carried out in expensive cleanroom environments. In this work, we propose and experimentally demonstrate a novel, single-step process for the rapid fabrication of high performance mid- and long-wave infrared EOT metasurfaces employing ultrafast direct laser writing (DLW). Microhole arrays composing extraordinary transmission metasurfaces were fabricated over areas of 4 mm² in timescales of units of minutes, employing single pulse ablation of 40 nm thick Au films on dielectric substrates mounted on a high-precision motorized stage. We show how by carefully characterizing the influence of only three key experimental parameters on the processed micro-morphologies (namely laser pulse energy, scan velocity and beam shaping slit), we can have on-demand control of the optical characteristics of the extraordinary transmission effect in terms of transmission wavelength, quality factor and polarization sensitivity of the resonances. To illustrate this concept, a set of EOT metasurfaces having different performances and operating in different spectral regimes has been successfully designed, fabricated and tested. Comparison between transmittance measurements and numerical simulations have revealed that all the fabricated devices behave as expected, thus demonstrating the high performance, flexibility and reliability of the proposed fabrication method. We believe that our findings provide the pillars for mass production of EOT metasurfaces with on-demand optical properties, and create new research trends towards single-step laser fabrication of metasurfaces with alternative geometries and/or functionalities.

1. Introduction

Due to a progressive miniaturization of technologies such as mobile phone cameras, optical circuits or sensing, optical metasurfaces are becoming one of the most promising key components towards the development of next generation lightweight devices. Metasurfaces consist of two-dimensional, subwavelength (typically resonant) building blocks, which can be either periodically or randomly arranged, and can be engineered to provide on-demand light control[1-8]. Contrary to conventional bulky optics, the ability to tailor light offered by metasurfaces does not rely on propagation effects (such as optical path length differences or linear absorption), but instead comes from abrupt amplitude and phase local or global discontinuities induced by localized and/or coupled resonances[1,2]. As a result, arrays of resonators can be specifically engineered to mimic, and even to outperform the functionalities of classical optics in a lightweight fashion[1-5,8]. Amongst the range of available metasurface functionalities, plasmonic-based metasurfaces exploiting the extraordinary optical transmission (EOT) effect have attracted much attention over the last years, as such devices are expected to play an important role in compact biological and chemical sensing[9-11], structural color generation or multispectral imaging[3,12,13]. The EOT phenomenon consists of a specific frequency band being transmitted through a periodically-arranged array of subwavelength nanoholes in an optically thin film. In particular, high transmission peaks appear close to the first Wood anomaly, i.e. at the frontier between the diffractive and sub-diffractive optical regimes[13–15]. The occurrence of this phenomenon has been studied in detail over the last years, and has been attributed to a resonant interaction between holes arranged in a lattice, assisted by highly confined fields associated to surface waves such as plasmon[15] or phonon polaritons[14,15]. The amount of transmission, the bandwidth, and spectral position of the EOT effect can be therefore controlled by design, via simply tuning the geometrical parameters (namely hole size, shape and periodicity) and/or constituent materials of the film, substrate and/or cover layer[8–10].

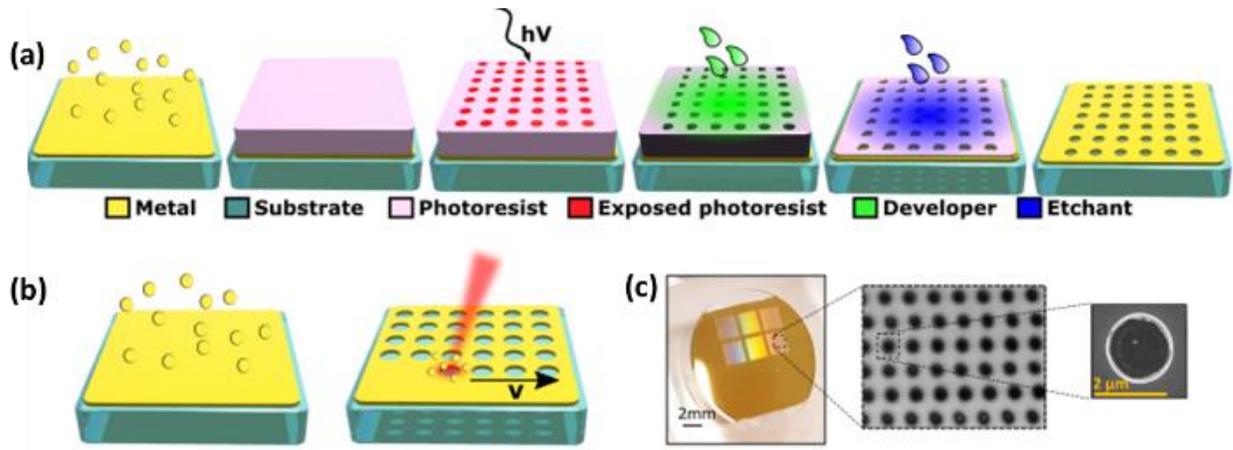

Figure 1. (a) Schematics of a typical lithographic process for micro- and nano- patterning of EOT metasurfaces, consisting of various fabrication steps. From left to right: after depositing a thin metal film, a photoresist layer is spin-coated on top of the surface, and exposed to electron or photo- lithography to modify the solubility of the exposed regions. Such regions are then removed employing a developer, followed by a wet/dry etching process to eliminate material from the exposed regions. The process ends by removing the photoresist/chemical leftovers, typically in acetone. (b) Alternative fabrication routine proposed in this work, consisting of direct laser patterning employing ultrafast scanned lasers. (c) Macroscopic view of one of the processed samples, with six 4 mm² areas patterned on it. Insets show an optical microscope and scanning electron microscope (SEM) image, confirming micrometric geometrical features.

While EOT metasurfaces introduce indubitable benefits to current optical and photonic technologies, their reliable large scale fabrication is currently at a low stage of maturity in terms of throughput and cost. This is primarily due to the small size of the patterned features and the high resolution required for the successful realization of metasurfaces (ranging from tens of nanometers to units of microns for electromagnetic metasurfaces operating from the ultraviolet to the mid-infrared spectral region, respectively)[1–3]. Their fabrication is currently achieved via established nanostructuring processes, on which an excellent overview can be found in Ref [16]. As generically depicted in Figure 1(a), amongst them, mask-based techniques such as electron-beam lithography (EBL) or photolithography combined with lift-off, nanoimprint, or wet/dry etching are nowadays the predominant fabrication methods chosen. Such techniques can indeed provide precise spatial resolution down to tens of nanometers[17–19], but at the same time require for expensive equipment operated under cleanroom environments[17–19]. An alternative technique with an extraordinarily high degree of versatility and throughput is based on ultrafast laser processing (ULP). This technique is arguably one of the most promising routes towards clean, large-scale and mass-production of nano- and micro-patterned devices, as it is based on single-step patterning procedures that do not require the use of polymer masks and contaminants, nor cleanroom equipment[16,20]. Employing sub-picosecond laser pulses, one can reduce the pulse energy required to trigger ablation, while achieving sharp contours in a wide variety of soft and brittle materials[20,21]. To date, a wide range of nano- and micro-features have been successfully fabricated using ultrafast lasers, including ripples and grooves, spikes, arrays of holes or pillars, hierarchical shapes or compositional and random structures[20]. Nevertheless, while ULP techniques have been now extensively explored towards the realization of e.g. high density data storage[22], biomimetic structures[20], photonic and microfluidic devices in transparent materials[23] or functionalized surfaces for wettability control[24,25], their potential towards fabrication of metasurfaces remains highly underexplored. Examples of recent works on ULP aiming at the production of metasurfaces include the fabrication of arrays of Si Mie resonators via direct interference laser pattering[26], arrays of Si nanoparticles via laser induced forward transfer[27], and arrays of microbumps in Au films using direct laser writing[28]. Additional work about randomly-structured metasurfaces for structural color generation and color encryption employing ultrafast lasers has been also recently reported[29,30].

In this paper, we introduce a novel fabrication process for the rapid realization of EOT metasurfaces in a single step, based on direct laser writing and thus free from expensive cleanroom environments. As depicted in Figure 1(b) and (c), this is achieved by focusing a pulsed laser at the surface of a thin metallic film deposited on a dielectric substrate, which is mounted on a motorized stage that moves in a zigzag fashion. By selecting the (previously calibrated) experimental setup parameters, a set of EOT devices with different optical and morphological characteristics has been designed, successfully fabricated over areas of 4 mm² in units of minutes, and its optical performance tested. Due to its rapid, cost-efficient and residue free nature, we believe that the herein proposed methodology puts EOT metasurfaces a step closer to industrial, real-world applications.

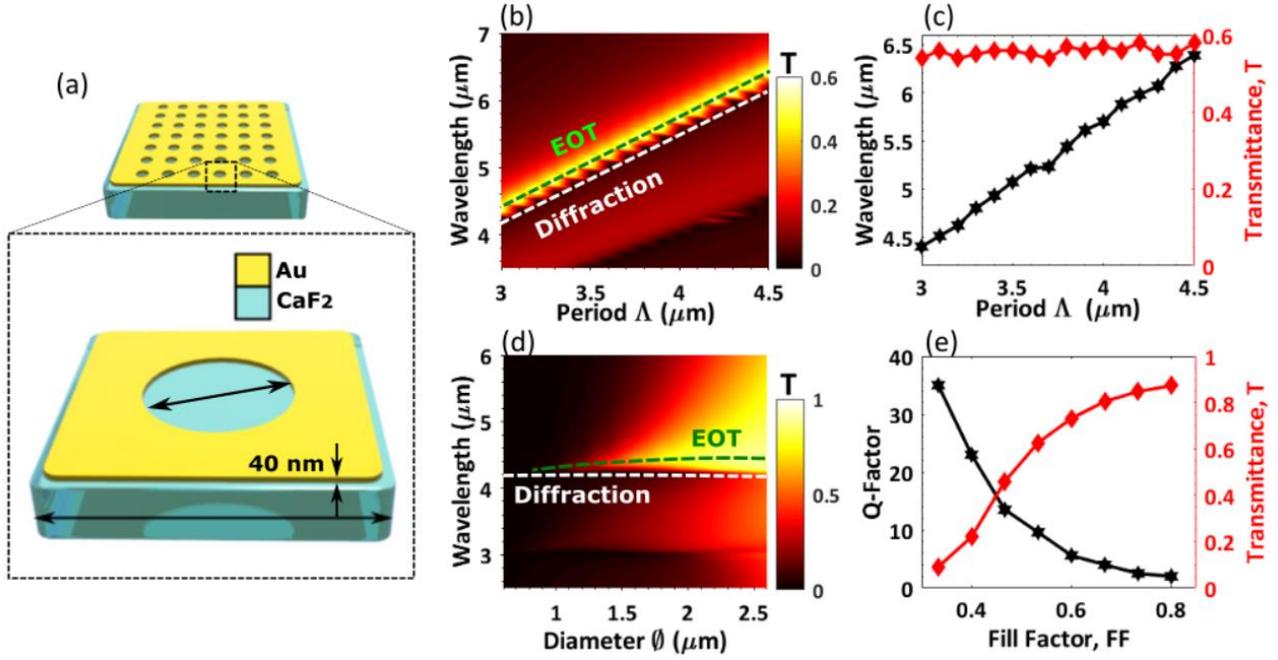

Figure 2. (a) Schematics of the plasmonic EOT metasurface unit cell, showing material distribution, dimensions and design variables (period $\Lambda$ and diameter $\emptyset$). (b) Transmittance spectrum (colorbar) as a function of the period, calculated imposing a constant fill factor of FF = 0.5. (c) Details of the variation of the EOT transmission peak wavelength (black) and transmittance (red) as a function of the period. (d) Transmittance spectrum (colorbar) as a function of the hole diameter, calculated while keeping the period at a constant value of $\Lambda$ = 3 µm. (e) Details of the variation of the Q-factor (black) and transmittance (red) as a function of the fill factor.

## 2. Methods

### 2.1 Finite element design and analysis

Prior to the identification and control of the key experimental laser parameters to be employed for the fabrication of EOT devices, it is necessary to understand the influence of the metasurface geometrical parameters on its optical response. For this purpose, we have carried out a finite element analysis using COMSOL Multiphysics ® (RF module). Technical details and boundary conditions of the simulation routine are discussed in the supplementary information, section S1.

A generic scheme of the metasurface and its unit cell considered in this work is depicted in Figure 2(a). The devices consist of a thin Au film (40 nm) on a CaF2 dielectric substrate. Au thickness was chosen to be thick enough to support high quality EOT, but thin enough to ensure full single-pulse perforation of the layer (i.e. slightly thicker than the linear optical penetration depth of gold (13 nm), which dominates absorption of ultrafast lasers in high extinction coefficient materials such as metals). CaF2 was selected as the substrate material due its high transparency across the near, mid- and long-wave infrared spectral regions[31]. The film is patterned with identical micro-holes arranged in a square periodic lattice. As shown in Figure 2(a), the proposed geometry gives two different geometrical variables, the period $\Lambda$ and the hole diameter $\emptyset$. Based on these two degrees of freedom, an additional parameter can be defined, namely the one-dimensional fill factor *FF*:

$$FF = \frac{\emptyset}{\Lambda} \qquad \text{eq. 1}$$

Figure 2(b) represents the computed transmittance spectrum of the proposed device as a function of the period, while imposing a fixed fill factor of FF = 0.5. As previously reported in literature[13,15], and confirmed by results in Figure 2(b), the EOT peak appears slightly red-shifted with respect to the frontier between the diffractive and sub-diffractive regimes. That is, when the first diffraction order $\theta_{m\pm1}$ becomes grazing to the surface, thus giving birth to the first Wood anomaly:

$$\lambda_0 = \Lambda n_{CaF_2} \qquad \text{eq. 2}$$

being $\lambda_0$ the free-space excitation wavelength and $n_{CaF_2}$ the refractive index of the substrate[31] ($n_{CaF_2}$ = 1.42 at a wavelength of $\lambda_0$ = 3 µm for reference). In Figure 2(c), we show

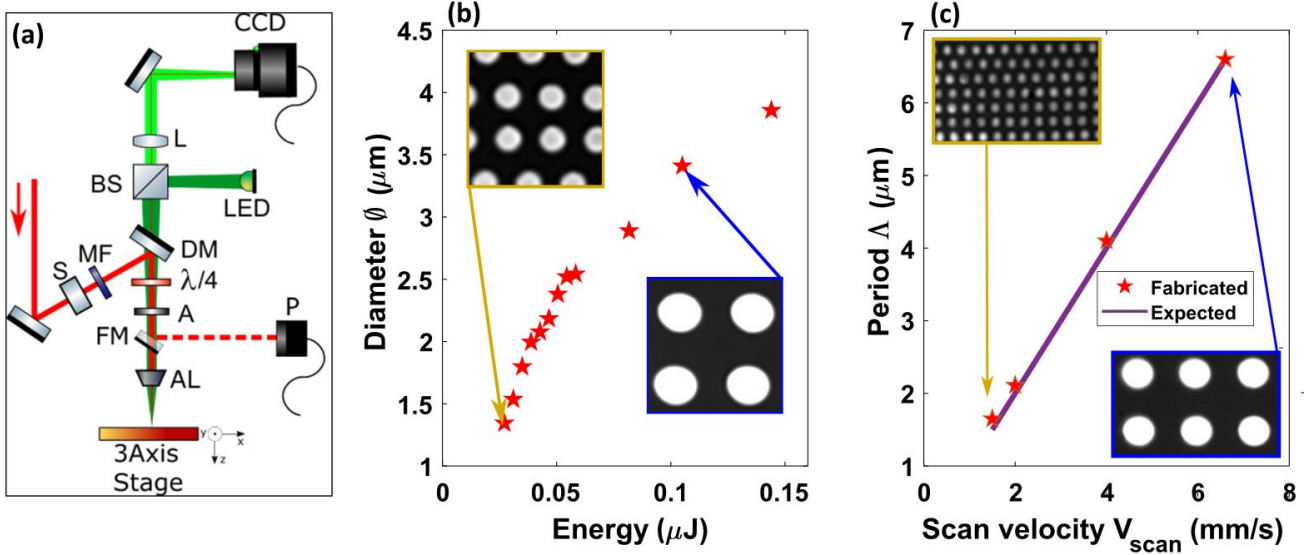

Figure 3 (a) Diagram showing the experimental setup used for the fabrication of EOT metasurfaces employing an ultrafast laser. Red lines correspond to the laser processing beam, whereas the green beam represents the in-situ observation path, illuminated by a LED. The labelled components are: slit (S), metallic filter (MF), dichroic mirror (DM), quarter-waveplate ($\lambda/4$), aperture (A), flip mirror (FM), aspheric lens (AL), beam splitter (BS), tube lens (L) and power meter (P). (b) Fabricated hole diameter vs. laser pulse energy for NA 0.25. The insets show two representative microscope images of the fabricated holes recorded in transmission mode, confirming the presence of light passing through the holes (field of view is ~12 μm x 12 μm). (c) Fabricated hole period vs. scaning speed for a repetition rate of $F_{rep} = 1$ kHz. The purple line represents the expected linear trend calculated via eq. 3, whereas the red symbols represent the experimentally measured periods fabricated. The insets show representative microscope transmission images of the fabricated hole arrays (field of view is ~18μm x 10μm).

fine details of the dependence of both, the EOT peak wavelength (black curve) and the amount of transmittance (red curve) with the geometrical period $\Lambda$. It can be seen that the EOT peak wavelength is almost linearly proportional to $\Lambda$, which is consistent with eq. 2. On the other hand, the amount of transmittance remains at a nearly stationary value of T ~ 0.57, which is associated to the use of a constant value $FF = 0.5$, while varying the geometrical period in our simulations.

In Figure 2(d), we show the transmittance spectrum as a function of the hole diameter for a fixed geometrical period of $\Lambda = 3$ μm, which implies a variation of the fill factor defined in eq. 1. Here, contrary to the previous study, both the diffraction frontier and the EOT peak remain nearly stationary, which is also consistent with eq. 2, since $\Lambda$ is fixed. Details of this simulation are revealed in Figure 2(e), where we represent the influence of the fill factor on the EOT quality factor (Q-factor = $\lambda_{peak}/\Delta\lambda$, black curve) and on the amount of transmittance (red curve). As it can be seen from the plot, the quality factor experiences an exponential-like decay when FF increases, accompanied by a logarithmic-like increase in transmittance. This behavior reveals a tradeoff between bandwidth of the EOT effect and the amount of light transmitted through the array.

### 2.2 Fabrication methodology and calibration of processing parameters

The finite element analysis discussed in the previous section demonstrates that the three main characteristics of the EOT effect, (spectral position of the EOT peak, transmittance and Q-factor) can be controlled by two key geometrical parameters, namely the period $\Lambda$ and the hole diameter $\emptyset$ (thus the fill factor FF), respectively. Therefore, our fabrication method has been specifically developed and calibrated to provide a flexible control of these geometrical variables. The experimental setup employed for the fabrication of EOT devices is depicted in Figure 3(a), and is conceptually similar to optical setups which have been successfully used for the writing of embedded optical waveguides described in [32]. It consists of two optical beam paths, separated by a dichroic mirror:

- Sample processing beam path fed by an ultrafast laser ($\lambda_0 = 1030$ nm, $\tau_{pulse} = 340$ fs, maximum repetition rate $F_{rep} = 2$ MHz, Gaussian profile), which is focused at the sample surface using an aspheric lens (AL).
- In-situ observation line employing a CCD camera and LED illumination, which is used for alignment purposes and in-situ monitoring of the fabrication process.

The laser pulse energy is controlled by a combination of motorized half-waveplate and thin film polarizer (not shown). A 45° flip mirror is placed prior to the aspheric lens in order to calibrate the incident laser pulse energy with a power meter P. The sample is mounted on a 3-axis stage, with the x-stage being an air-bearing high-speed, high-precision translation stage with scan velocities up to $V_{scan} = 20$ mm/s. The samples used were 40 nm thick Au films evaporated on CaF2 substrates (fabrication of the films is described in the supplementary information, section S2). It should be noted that the laser irradiation plane (x-z) was horizontal and the sample mounted vertical, in order to avoid re-deposition of laser-ablated material that could contaminate the sample.

The fabricated hole diameter was controlled via adjusting the energy of the focused laser pulse at the sample surface. In order to increase the range of attainable diameters, two different objective lenses with different numerical apertures have been used (NA1 = 0.25 and NA2 = 0.47). Results of the influence of the pulse energy on the hole diameter ∅ for NA1 = 0.25 are displayed in Figure 3(b), showing a predictive, increasing behavior with a bending at very high energies. This curve bending is a consequence of the Gaussian-like intensity distribution of the beam at the focal plane, further discussed in section S3.1 from the supplementary material [33,34]. This predictive and reproducible behavior allows one to calculate the output energy required for the fabrication of a specific hole diameter. An equivalent calibration curve of hole diameter versus pulse energy was obtained also for NA = 0.47, as also detailed in section S3.1 from the supplementary information. Importantly, the study of the influence of the energy on the diameter of the ablated spot was complemented with AFM topography measurements (see supplementary material section S3.2), which revealed clean film ablation with sharp borders for both lenses, confirming that the pre-selected Au thickness (40 nm) was optimum for our experimental conditions.

The desired geometrical period Λ for a given metasurface was imprinted by high-speed scanning of the sample along the x-axis in a zigzag fashion, where holes were fabricated via single pulse ablation (i.e. row by row). Periods in the x direction (Λx) were adjusted via modification of the scan velocity $V_{scan}$ for a fixed $F_{rep}$, according to

$$\Lambda = \frac{V_{scan}}{F_{rep}} \qquad \text{eq. 3}$$

whereas periods in the y direction (Λy) were adjusted via discrete displacements of the y-motor after writing of each row.

Using this strategy, the maximum achievable processing speed is ultimately limited by the pulse repetition rate of the laser, as well as by the sample scan velocity. All the devices reported in this work have been fabricated for a fixed repetition rate of $F_{rep}$ = 1 kHz and different geometrical periods were achieved via varying $V_{scan}$. The purple line shown in Figure 3(c) corresponds to the prediction of the fabricated period Λ by eq. 3. Such a linear relation was indeed experimentally verified over the entire range of periods studied (Λ = 1.6 μm – 6.6 μm), as confirmed by the red data points in Figure 3(c). As also appreciated in the insets of Figure 3(c), occasional stitching errors occur in some of the processed areas as a consequence of small delays of the motorized stage after writing each row (thus minor laser/stage synchronization mismatches). This affects the separation of holes written in different rows but not the separation of holes within the same row. As we will see in the results section, this was found to have only minor effects on the optical response of our devices, and could be minimized employing multiplexed beams for processing multiple holes in a single scan[35] or even eliminated by an improved synchronization between translation stage and laser repetition rate.

3. **Results**

A range of different EOT metasurfaces with areas of 2 mm x 2 mm have been fabricated using the method described above, i.e. selecting the required NA, pulse energies and scan velocity. The results presented below show their different optical response in terms of quality factor (section 3.1), optical transmission peak (section 3.2), and polarization-selective extraordinary transmission (section 3.3).

3.1 **EOT devices with different quality factors**.

According to the simulations shown in Figure 2(d) and 2(e), the Q-factor and amount of transmittance of the EOT device can be controlled via increasing the hole diameter

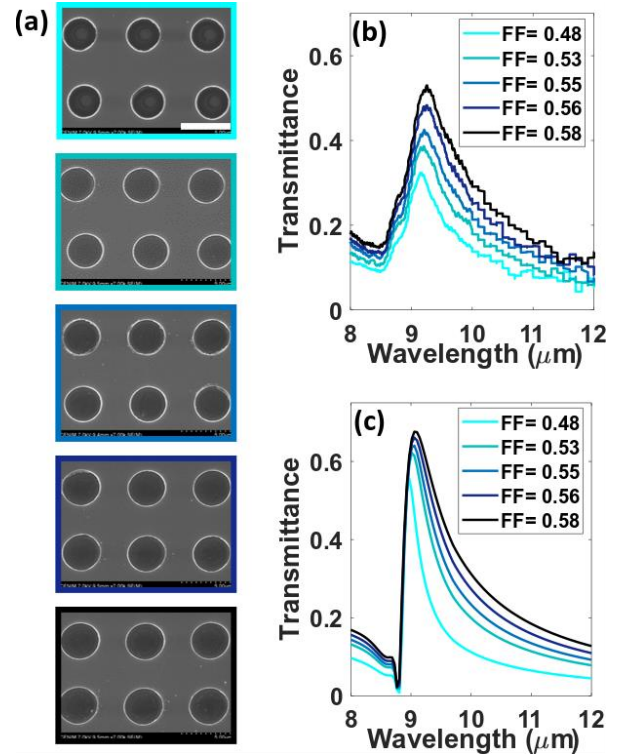

Figure 4. (a) SEM images of five different laser-fabricated EOT metasurface devices with different fill factors and identical geometrical period of Λ = 6.6 μm (scale bar is 5μm). (b) FTIR measurements of the fabricated samples, revealing a progressive increase of the transmittance as the fill factor increases, in line with the predictions by numerical simulations shown in (c).

Ø for a fixed period Λ, and therefore via increasing the fill factor *FF*. In order to demonstrate this concept experimentally and highlight the flexibility of the laser processing approach, five different arrays with and different hole diameters but identical periods have been fabricated. As detailed in table S1 from the supplementary information (section S4), all devices were processed using $V_{scan}$ = 6.6 mm/s and $F_{rep}$ = 1 kHz, yielding a period of Λ = 6.6 μm. Different pulse energies were used in order to adjust the hole diameters, and each device took 3 minutes to be fabricated. SEM images of as-fabricated devices, with nominal fill factors of FF = 0.48, FF = 0.53, FF = 0.55, FF = 0.56 and FF = 0.58 are displayed in Figure 4(a). Transmittance spectra of these devices were then measured using Fourier transform infrared spectroscopy (FTIR). Results are shown in Figure 4(b), where an increase in transmittance of the EOT peak with subsequent decrease in quality factor can be clearly observed as the fill factor increases. Since the period of all devices was fixed, the transmission peak wavelength remains at a nearly stationary value ($\lambda_0$~ 9.2 μm). To compare our results with theoretical EOT performances, as-fabricated dimensions of all the devices were introduced in COMSOL, and transmittance spectra were calculated yielding the results shown in Figure 4(c). Indeed, a very good agreement with the experimental results was obtained, confirming the same overall shape of the transmission spectra, peak position and spectral evolution of the measured devices (Figure 4(b)) when increasing the fill factor.

### 3.2 EOT devices with on-demand transmission peaks.

The capability of our fabrication method to generate extraordinary transmission at specific on-demand wavelength peaks was also explored. As shown in the previous sections (Figures 2(b) and 2(c)), this can be achieved via adjusting the period Λ between holes. To this end, a new set of devices having different periods was fabricated and tested. As shown in table S2 (supplementary information section S4), the scanning speed was varied for a fixed repetition rate of 1 kHz, yielding as-fabricated periods of Λ=2.1 μm ($V_{scan}$ = 2.0 mm/s), Λ=4.1 μm (Vscan = 4.0 mm/s) and Λ=6.6 μm ($V_{scan}$ = 6.6 mm/s), with total processing times of 19 min, 10 min, and 3 min respectively. The pulse energy was adjusted accordingly to create spot sizes with an approximate fill factor F = 0.5. For the lowest speed/smallest hole diameter, the high numerical aperture lens (NA = 0.47) was used (see energy vs hole diameter calibration in Figure S2(b) from the supplementary information). SEM images of the devices are shown in Figure 5(a), and their transmittance spectra were measured employing FTIR. As shown in Figure 5(b), the EOT peak shifts towards longer wavelengths as the period increases. Numerical simulations of the optical response, using the period and hole diameters determined from the SEM images, are shown in Figure 5(c). A good agreement with the experimental results is found also here. In particular, the peak positions and spectral widths predicted match well those of the measured devices. Moreover, the secondary transmission peak in the diffraction regime of each device, predicted by the calculations, is also present in our measurements. Finally, a general decrease of maximum transmittance (ΔT ~ 0.15) is observed in experiments with respect to simulations. This might be due to imperfect hole shapes, small stitching errors and roughness of the Au film itself (not taken into account in simulations). In particular, in section S5 of the supplementary information we explicitly show (via FEM simulations) how imperfections of hole shapes have only minor effects on both the charge distribution of the metasurface when in resonance, as well as on its amount of transmittance.

### 3.3 EOT devices with polarization-selective response.

Finally, the potential of our laser processing method to induce anisotropic optical responses for polarization selective extraordinary transmission was also investigated. As depicted in Figure 6(a), square lattices were replaced by rectangular ones, giving rise to two different periods Λx and Λy. Moreover, circular holes were replaced by elliptical ones, with different diameters in the x (Øx) and y (Øx) direction (Figure 6(b). The required elliptical focal spots were achieved by inserting a slit aligned along the x-axis)

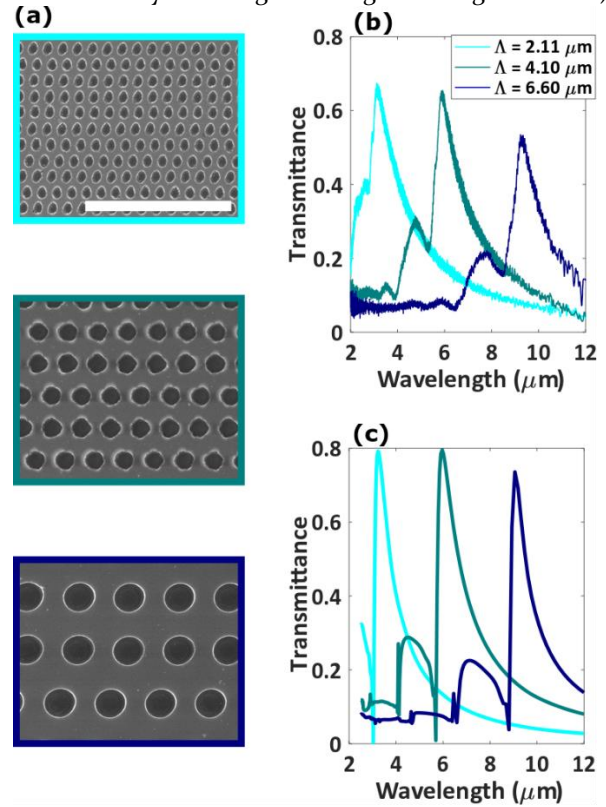

Figure 5. (a) SEM pictures of three laser-fabricated EOT metasurface devices with increasing geometrical periods (scalebar is 10 μm). (b) FTIR measurements of the fabricated samples, revealing a progressive increase of the transmission peak wavelength as the period increases. The results are in very good agreement with the numerical simulations shown in (c).

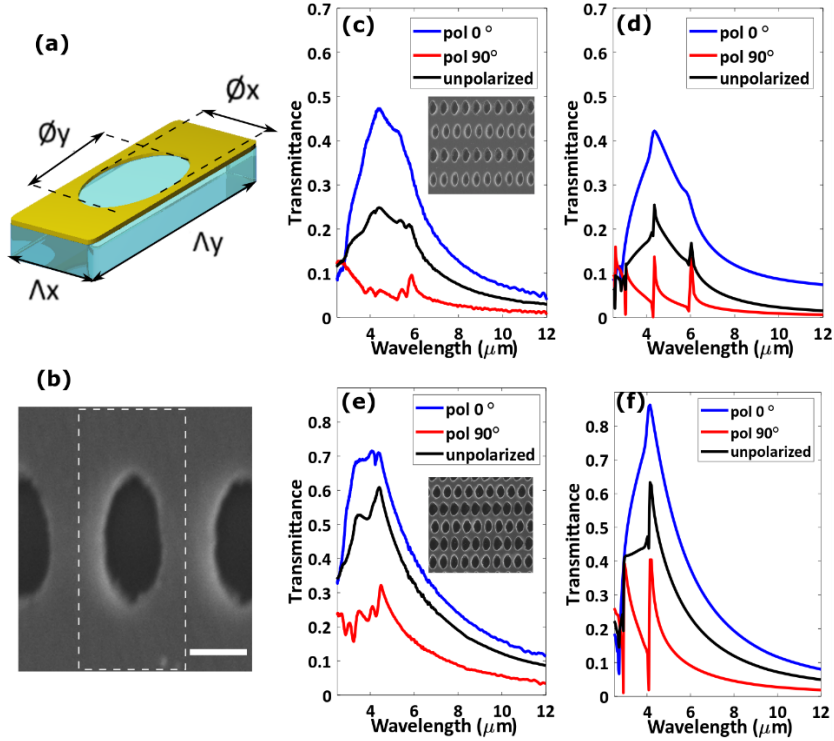

Figure 6 (a) Unit cell of EOT metasurfaces with polarization-selective optical response, where both, period and diameters can be modified in two dimensions (x and y axes). (b) High magnification SEM image revealing fine details of as-fabricated elliptical unit cells (scale bar is 1 µm). (c-d) Measured (c) and simulated (d) polarization selective transmittance spectra for device E1. The inset shows a SEM image of the device (field of view is 20 µm x 18 µm). (e-f) Measured (e) and simulated (f) polarization selective transmittance spectra for device E2. Inset shows an SEM image of the device (same field of view as in (c)).

in the optical path of the laser processing line, as shown in Figure 3(a). That way, the entrance pupil of the focusing aspheric lens (AL) is underfilled in the y-axis, leading to a reduced focusing and thus diameter increase along the y-axis. For a complete description of the influence of the slit on the spot size, the modified beam divergence in the y-direction, leading to astigmatism, also needs to be taken into account[36].

As a proof of concept, two different arrays of elliptical holes in rectangular lattices (namely E1 and E2) were fabricated. Table S3 from the supplementary information shows the laser parameters employed for their fabrication, as well as their as-fabricated dimensions. The period in the x-axis was fixed to $\Lambda_x$ = 2 µm in both devices. $\emptyset_x$ was chosen to be larger and $\Lambda_y$ smaller for E2 than for E1, in order to fabricate devices with different fill factors (and thus different amount of transmittance). Each device took 8 minutes to be fabricated. The measured polarization-dependent transmission spectrum of device E1 is shown in Figure 6(c), revealing a peak transmission for light polarized at 0° of T ≈ 0.5 at a wavelength of $\lambda_0$ ~ 4.2 µm, and transmittance T < 0.1 for polarization at 90°. This strong polarization anisotropy was corroborated by the results of numerical simulations shown in Figure 6(d), showing excellent agreement. It is worth noting that the predicted higher-order Wood anomalies possessing high Q-factors for 90° polarization are much weaker in the experimental results. This can be considered as a beneficial side effect of the roughness of the films well as of the small fluctuations in the hole diameter and lattice order achieved with the laser fabrication method. The presence of such small imperfections results, in this case, in an improved contrast ratio between linear polarization states (i.e. $\Delta T = T(0°)/T(90°)$) when compared to the simulation.

The experimental transmittance spectrum for device E2 is displayed in Figure 6(e), showing a similar behavior as device E1 but with a higher fill factor and narrower resonance. Both, the transmission peaks for polarization at 0° and 90° are higher than for E1, essentially due to a larger elliptical spot size (and smaller period in y) chosen, which is also confirmed by calculations in Figure 6(f). Also here, the predicted high-frequency resonances are efficiently attenuated.

## 4. Conclusions

We have demonstrated a versatile processing technique based on ultrafast direct laser writing for the reliable fabrication of high performance EOT metasurfaces. Contrary to lithographic-based fabrication methodologies, our technique enables the single-step realization of EOT devices of several mm[2] in a few minutes, carried out in cleanroom-free environments without generating chemical residues. It was shown that the main performance characteristics of EOT devices in terms of transmission wavelength, amount of transmittance and Q-factor can be controlled deliberately via tuning of the period $\Lambda$, the hole diameter $\emptyset$ and the hole ellipticity by means of adjustment of the sample scanning speed, laser pulse energy and shape of the spot respectively. These calibrations have then been successfully employed to fabricate various sets of devices with on-demand control of the transmission peak, Q factor and polarization sensitivity. FTIR measurements have revealed striking agreement between as-fabricated and simulated devices, thus demonstrating the validity, flexibility and versatility of our approach.

Based on the reported proof-of-concepts, it appears straight-forward to fabricate cm² areas in the same times reported in this paper, by means of exploiting the high-repetition rate of modern ultrafast laser systems, and/or multiplexing techniques. The variety of fabricated shapes can be extended to more complex geometries using alternative beams (such as vortex beams for the fabrication of coaxial apertures)[37] based on the use of waveplates or spatial light modulators. Diameters down to 40 nm (i.e. far below the diffraction limit) have been reported in direct laser writing processes[38], thus both fabrication of deep-subwavelength structures and scalability of our approach towards shorter wavelengths (e.g. visible and near IR spectra) is also possible.

In conclusion, we believe the proposed methodology could be applied not only to other metasurface geometries, or materials, but also for the rapid creation of large area molds for nanoimprint lithography, nanocasting, and stencils or masks for lift off processes[39]. Fabrication of more complex unit cells for alternative metasurface functionalities (such as multipixel meta-atoms containing different geometrical sizes for the fast processing of e.g. beam steering devices[6,40]) could be carried out via the use of multiple laser scans with different energies (thus different sizes of the ablation area).

## AUTHOR INFORMATION

### Corresponding Authors
* cr408@io.cfmac.csic.es, j.siegel@csic.es

### Author Contributions
The manuscript was written through contributions of all authors. ‡These authors contributed equally.

### Funding Sources
This work was financially supported by the Spanish Research Agency (MCIU/AEI/Spain) through project TEC2017-82464-R and PID2020-112770RB-C21, as well as by the National Research Council of Spain (CSIC) for the intramurales project 201850E057.

## ACKNOWLEDGMENT
Authors would like to thank Ms. Hannah Barnard and Prof. Geoffrey Nash (University of Exeter) for providing polarization dependent FTIR measurements.

## ABBREVIATIONS
AFM, Atomic force Microscopy, EOT, Extraordinary Optical Transmission; FTIR, Fourier Transform Infrared Spectroscopy; FF, Fill Factor; SEM, scanning Electron Microsocopy; EBL, electron-beam lithography; ULP, ultrafast laser processing; DLW, direct laser writing

## SUPPORTING INFORMATION
S1. Finite element model for EOT metasurfaces. S2. Fabrication of Au films on CaF2 substrates. S3. Dependence of the hole diameter with laser pulse energy. S3.1. Topography of the ablated region and idealness of the Au film thickness. S4. Laser processing parameters and as-fabricated dimensions of EOT metasurfaces. S5. Influence of the hole-shape imperfections on the optical response.

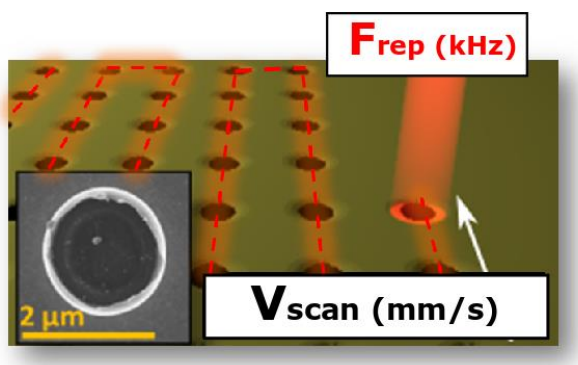
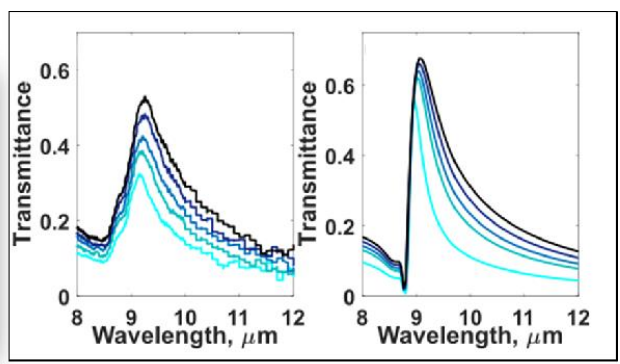